\documentclass[runningheads]{llncs}
\usepackage{graphicx}
\usepackage{subfigure}
\begin{document}
\title{Reproducible White Matter Tract Segmentation Using 3D U-Net on a Large-scale DTI Dataset}
\titlerunning{Direct and Reproducible White Matter Tract Segmentation}
% If the paper title is too long for the running head, you can set
% an abbreviated paper title here
%
\author{Bo Li\inst{1,2}\and
Marius de Groot\inst{2} \and
Meike W. Vernooij\inst{2} \and
M. Arfan Ikram\inst{2}\and
Wiro J. Niessen\inst{2,3}\and
Esther E. Bron\inst{2}} 
\authorrunning{B. Li et al.}
% First names are abbreviated in the running head.
% If there are more than two authors, 'et al.' is used.

\institute{Northeastern University, Shenyang, China 
\and
Erasmus MC, Rotterdam, The Netherlands
\and
Delft University of Technology, Delft, the Netherlands
\\
\email{\{b.li, e.bron\}@erasmusmc.nl}}

\maketitle          % typeset the header of the contribution
\begin{abstract}Tract-specific diffusion measures, as derived from brain diffusion MRI, have been linked to white matter tract structural integrity and neurodegeneration. As a consequence, there is a large interest in the automatic segmentation of white matter tract in diffusion tensor MRI data. Methods based on the tractography are popular for white matter tract segmentation. However, because of the limited consistency and long processing time, such methods may not be suitable for clinical practice. We therefore developed a novel convolutional neural network based method to directly segment white matter tract trained on a low-resolution dataset of 9149 DTI images. The method is optimized on input, loss function and network architecture selections. We evaluated both segmentation accuracy and reproducibility, and reproducibility of determining tract-specific diffusion measures. The reproducibility of the method is higher than that of the reference standard and the determined diffusion measures are consistent. Therefore, we expect our method to be applicable in clinical practice and in longitudinal analysis of white matter microstructure.
%Our method achieved a dice coefficient of 0.66 for forceps minor (FMI) segmentation and 0.77 for right corticospinal tract (CST). 

% The change in white matter tract plays such an important role in neurodegeneration that considerable effort is put into improving the segmentation which enables its quantification.

% and determine its applicability for the analysis of white matter microstructure in neurodegeneration, 

\end{abstract}
\section{Introduction}
% Neurodegenerative disease is increasingly prevalent mainly due to the aging population worldwide and is leading to an enormous burden on many individuals as well as society as a whole [1]. Therefore, it is important to improve understanding of neurodegenerative disease. In particular, changes in white matter (WM) tracts, which are the neural fibers enabling communication among brain regions. Changes in WM tracts have increasingly been associated with cognitive dysfunction and neurodegeneration. To quantitatively describe the change, a precise segmentation method is needed [2, 3].

White matter (WM) tracts are the neural fibers enabling communication among brain regions. The changes in which have increasingly been associated with cognitive dysfunction and neurodegeneration. For improving understanding of neurodegenerative process and the study of pathogenesis triggered by abnormal changes, a quantitative description of WM tract is essential. Therefore, a precise segmentation method used for quantifying WM tract is needed \cite{o2001evidence}. 
%Different strategies of step size, angular constraint, interpolation and seeding give different results of tractography.

Tract segmentation is typically performed by tractography followed by a filtering step based on the prior information. After tractography reconstruction, millions of possible pathways are filtered into specific tract either via tract-specific thresholds \cite{de2015tract}, anatomical atlas based mask \cite{lawes2008atlas,o2007automatic} or neighboring anatomical labels based prior probability \cite{yendiki2016joint}. However, these steps result in accumulating intermediate errors, multiple environment settings and limited consistency due to the property of tractography and, therefore, limit their application in clinical practice.

The U-Net architecture \cite{ronneberger2015u} has shown good performances in several segmentation tasks. Based on 3D U-Net, the newer V-Net \cite{milletari2016v} made further improvements by introducing residual function, strided convolution and convolution transpose operations. Recently a U-Net based WM tract segmentation method \cite{wasserthal2017direct} showed competitive results to tractography-based methods. Model in \cite{wasserthal2017direct} was trained on a high resolution dataset of only 20 subjects. In this paper we develop a method based on a large dataset of lower resolution data, and evaluate the potential of the method in this setting.

% However, we expect a deep learning method to especially have high potential in the opposite case of a large-scale dataset of low resolution.
% potential for direct WM tracts segmentation. 
% This work presents a novel method for direct segmentation of white matter tracts on a large-scale dataset.

This work presents a novel deep learning method for direct segmentation of white matter tract. Our method was evaluated on the tasks of FMI and CST segmentation and determining diffusion measures. We will evaluate whether this method is reproducible and can be used to provide more insight into the role of WM microstructure in neurodegeneration. 

\section{Methods}
\subsection{Model}
We built our model based on the 3D U-Net architecture. We added batch normalization after each convolution layer and replaced Relu activation function with PRelu. The used convolution layers are 3D with a kernel size of 3$\times$3$\times$3.

As input to the model, voxel-wise diffusion tensor elements were used. The input was fed in random batches during each training iteration to increase the robustness. Its batch generation was ``on-the-fly'' paralleled to the training process for efficiency. The method outputs a binary segmentation of a specific tract.
% The generation of input batch was ``on-the-fly'' paralleled to the training process for efficiency.

\subsection{Dataset}
% The method was developed based on Rotterdam Study (RS) dataset. This is an ongoing, prospective, population based cohort study targeting causes and consequences of age-related diseases among 14,926 subjects aged 45 years or over [11]. The RS was approved by the medical ethics committee. Written informed consent was obtained from all participants. After quality assessment, 9149 MRI scans from 4983 non-demented elderly subjects were available for this work. 
% The method using the dataset of Rotterdam Study (RS), an ongoing, population based cohort study [11]. After quality assessment, 9149 MRI scans from 4983 non-demented elderly subjects (age: 45+ years) were available for this work. Scans were performed at 1.5 Tesla (GE Signa Excite). The diffusion weighted images (DWIs) were acquired with a maximum b-value of 1000s/mm$^2$ in 25 gradient directions. Voxel size was resampled from 2.2 x 3.3 x 3.5 to 1 mm$^3$.

The method was developed based on the dataset of Rotterdam Study, an ongoing, population based cohort study \cite{hofman2015rotterdam}. After quality assessment, 9149 MRI scans from 4983 non-demented subjects were available for this work. Scans were performed at 1.5 Tesla. The diffusion weighted images (DWIs) were acquired with a maximum b-value of 1000 s/mm$^2$ in 25 gradient directions. Voxel size was resampled from 2.2$\times$3.3$\times$3.5 mm$^3$ to 1 mm$^3$.

We assign these scans into an optimization set (D1), a validation set (D2) and a reproducibility set (D3). Their sizes are as follows: $D1a_{train}$ 864 subjects, $D1a_{test}$ 218 subjects, size is same for $D1b_{train}$ and $D1b_{test}$ but with different subjects; $D2_{train}$ 7162 scans (including D1), $D2_{validate}$ 200 subjects and $D2_{test}$ 1036 subjects; $D3_{test}$ 80 subjects. The subjects (mean age of 69.7 years) in $D3_{test}$ had been scanned twice (mean interval of 19.3 days). A separate cohort was used for $D2_{validate}$ and $D2_{test}$ to ensure this is completely independent from $D2_{train}$. Additionally, all $D3_{test}$ related scans, which are their other rounds of scan, were excluded from D2 for the purpose of reproducibility evaluation.
%As most subjects have multiple scans, 
\subsection{Preprocessing}
% DWIs were corrected motion and eddy currents by co-registering all diffusion weighted volumes to the $b = 0$ volume with Elastix [11]. Diffusion tensors were estimated with a Levenberg–Marquard nonlinear least squares optimization algorithm, as available in ExploreDTI [12]. Diffusion measurements, such as fractional anisotropy (FA) and mean diffusivity (MD), were computed based on the estimated diffusion tensors. 

DWIs were corrected motion and eddy currents by co-registering all diffusion weighted volumes to the $b = 0$ volume with Elastix \cite{klein2010elastix}. Diffusion tensors were estimated with ExploreDTI \cite{leemans2009exploredti}. Diffusion measures, such as fractional anisotropy (FA) and mean diffusivity (MD), were computed based on the estimated tensors. 
% Diffusion tensors were estimated with a Levenberg–Marquard nonlinear least squares optimization algorithm, as available in ExploreDTI [12]. Diffusion measurements, such as fractional anisotropy (FA) and mean diffusivity (MD), were computed based on the estimated diffusion tensors. 

The diffusion tensor imaging (DTI) was used because this was the most suitable model for low-resolution DWIs. To evaluate location and structure information, with FLIRT \cite{jenkinson2002improved} we registered the MNI\_152 template and T1 weighted image (T1) to DTI space where most features were computed. Tissue masks including WM and gray matter (GM) were applied on all features. Due to the large image size and computation limitation of 3D convolution, we computed the region of interest (ROI) as input based on tract bounding boxes. The ROI sizes are 96$\times$64$\times$64 (FMI) and 64$\times$96$\times$128 (CST).
% So voxel-wise location in MNI (location) is also available for feature selection. 
\subsection{Reference standard}
% As reference standard we used a clinical-accepted method [3] achieved by probabilistic tractography followed tract-specific thresholds.
As reference standard we used a clinical-accepted method \cite{de2015tract}, which consists of probabilistic tractography and tract-specific thresholds. Manual annotation can not be obtained as WM tracts are not visible on imaging and the semi-manual annotation on tractography images is also unrealistic for such a large dataset. 

The method was evaluated on FMI and CST tract, since they are significantly related to aging \cite{de2015tract}, anatomically distinctive and have different degrees of difficulty for segmentation \cite{wasserthal2017direct}. 

% The method was evaluated on FMI and CST tract, as these tracts have clinical value because of their significant relation with aging [4]. Moreover, these two tracts are anatomically distinctive and have different degrees of difficulty for segmentation [8]. CST belongs to ``hard" group of segmentation and FMI is much smaller than CST.

\subsection{Evaluation metrics}
Segmentation accuracy was quantified by the Dice coefficient (DC). Binary segmentations were created from the probabilistic output by thresholding by 0.5.

To evaluate reproducibility, tract-specific metrics were compared between our method and the reference standard. Median FA and MD were individually computed inside the segmented tract, then averaged over $D3_{test}$. We computed the R$^2$ value of ordinary least squares (OLS) regression for measures in both scans. Cohen's kappa ($K$), which measures inter-rater agreement, was computed by rigidly aligning the FA image of rescan to the space of the first scan. We used t-test to compare $K$ and paired scan-rescan differences of FA, MD and volumes with those of the reference standard, and used paired t-test to compare whether the measures determined by our method are consistent in both scans. 

% \subsection{System}
% The experiments were ran on one node of Cartesius, Dutch national supercomputer. Such a node consists of an Intel E5-2450 v2 CPU working at 2.5 GHz with 96 GB of memory and two NVidia Tesla K40m GPUs with 11 GB of memory.

\section{Experiments}
% The experiments were ran on one node of Cartesius, Dutch national supercomputer. Such a node consists of an Intel E5-2450 v2 CPU working at 2.5 GHz with 96 GB of memory and two NVidia Tesla K40m GPUs with 11 GB of memory.
The experiments were ran on one node of Cartesius, Dutch national supercomputer, with the Intel E5-2450 v2 CPU and NVidia Tesla K40m GPU.
\subsection{Method optimization experiments}
We optimized the method using the FMI tract on three key elements: 1) input, 2) the loss function and tract weight, and 3) network architecture. Experiments 1) and 3) were performed on $D1a$ and $D1b$, trained with default parameters of optimizers; experiment 2) was performed on $D2$. The following paragraphs will describe these optimization experiments.

We trained the V-Net based model with eleven different inputs. Nadam optimizer \cite{dozat2016incorporating} and weighted inner product \cite{choi2010survey} loss function ($L_{wip}$) were used. The choice of input is based on DC and computation consumption. Since there are 25 diffusion weighted volumes in our raw DWIs and the number of volumes increases with resolution, e.g. 270 volumes in 7T scanner, it's an essential step to choose a concentrated and generalized input that works on different scanners. We considered diffusion tensor, FA, MD, location and T1 for training the model. we experimented to find the efficient input to avoid overlapping information and to reduce our high computation load due to the 3D convolution and large dataset.

%As some of them were expected to contain overlapping information, we experimented to find the optimal feature combination. Moreover, our high computation load due to the 3D convolution and large dataset can therefore be reduced by using efficient input.

% The choice of architecture is based on the test DC. 

To compare the $L_{wip}$ and weighted cross entropy ($L_{wce}$) loss function, we trained two V-Net based models using  
% \begin{equation}
%  L_{dice} = - \frac{2* \sum_{i=1}^{N} r_i * p_i * W}{\sum_{i=1}^{N} r_i + \sum_{i=1}^{N} p_i} 
% \end{equation}
 \begin{equation}
 L_{wip} = - \frac{1}{N}\sum_{i=1}^{N}  W *r_i * p_{i} + (1 - r_i) * (1 - p_{i}) 
\end{equation}
 and 
 \begin{equation}
 L_{wce} = - \frac{1}{N}\sum_{i=1}^{N} W *r_i * log(p_{i}) + (1 - r_i) * log(1 - p_{i}), 
\end{equation}
where $ r_i \rightarrow \{0, 1\} $ is the reference standard, $ p_i \rightarrow \{0, 1\} $ is the binarized prediction, $N$ is the voxel number of the input and $ W=[1, 3, 5, 10, 100] $ is the weight of tract. Due to the great frequency imbalance between classes, we evaluated different weights (W) for FMI segmentation, ranging from 1 to the mean frequency ratio of non-tract relative to tract, i.e. 100. Models were trained using Adam optimizer with an initial learning rate of 0.1, which was automatically reduced by $50\%$ once the validation loss stopped improving for 10 epochs.

Similarly, to investigate if the newer V-Net architecture performs better than 3D U-Net in WM tract segmentation, two separate models were trained using diffusion tensor input and $L_{wip}$. Furthermore, to avoid the chance that one gradient descent algorithm works better for a particular back-propagation pathway, we doubled the number of experiments using Adam \cite{kingma2014adam} and Nadam optimizer. 
 
% $L_{dice}$ and $L_{wce}$ are widely used loss functions for semantic segmentation tasks. $L_{dice}$ explicitly optimizes the method towards higher DC, while $L_{wce}$ also effects on the probabilistic distribution of non-tract background. 

\subsection{Validation experiments}
The optimized method was trained for FMI and CST tract on $D2_{train}$ to evaluate accuracy ($D2_{test}$) and reproducibility ($D3_{test}$). For $D3_{test}$, because of the short time interval between two scans (i.e. 19.3 days on average), tract segmentations and diffusion metrics are expected to be identical. We computed the paired scan-rescan differences, mean, standard deviation, $R^2$ value for FA, MD and volumes inside the segmented tracts in both scans and the Cohen's kappa to evaluate reproducibility of both segmentation and determining diffusion measures.

\section{Results}
\subsection{Method optimization results}

\begin{figure}[!ht]
  \centering
   \includegraphics[scale=0.46]{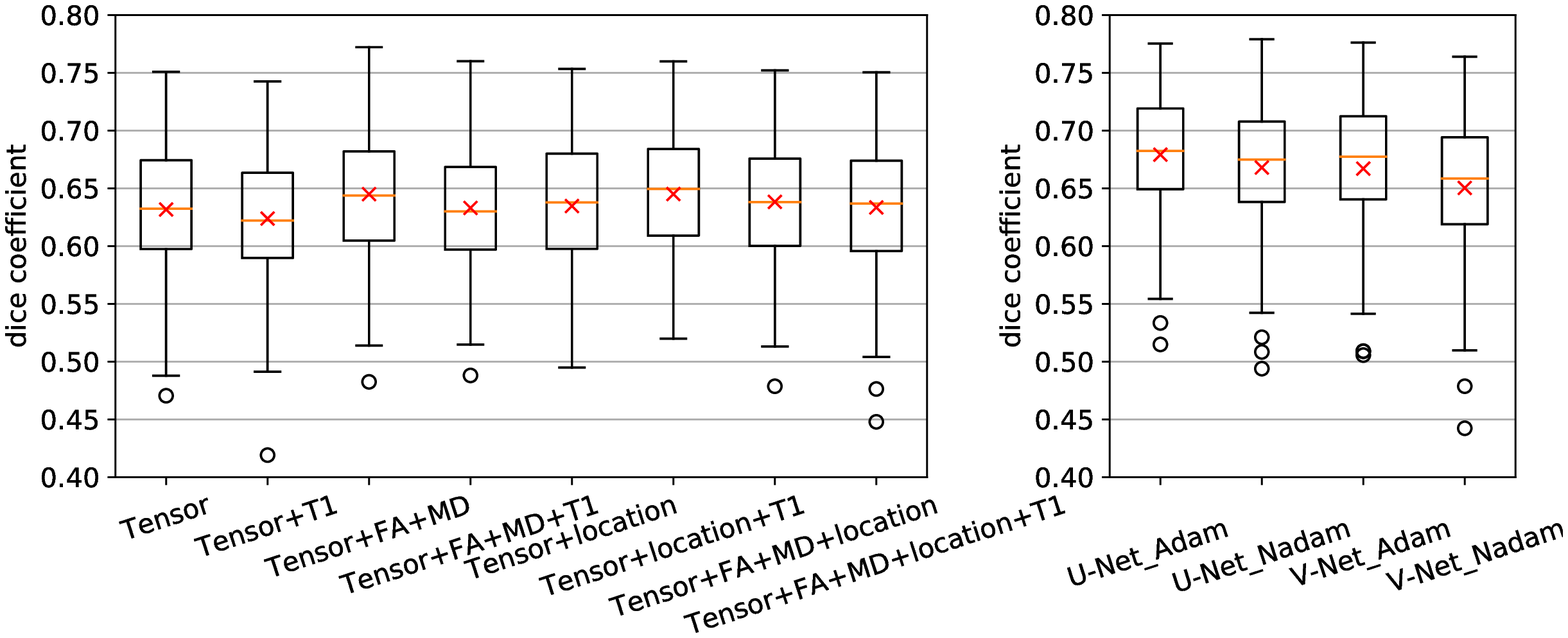}
  \caption{Test dice coefficient of FMI for different: (left). Input images. The ``Location'' is an image of voxel-wise coordinates on MNI\_152 template; (right). Architecture and optimizer.}\label{fig1}
    \vspace{0.3cm}
  \includegraphics[scale=0.46]{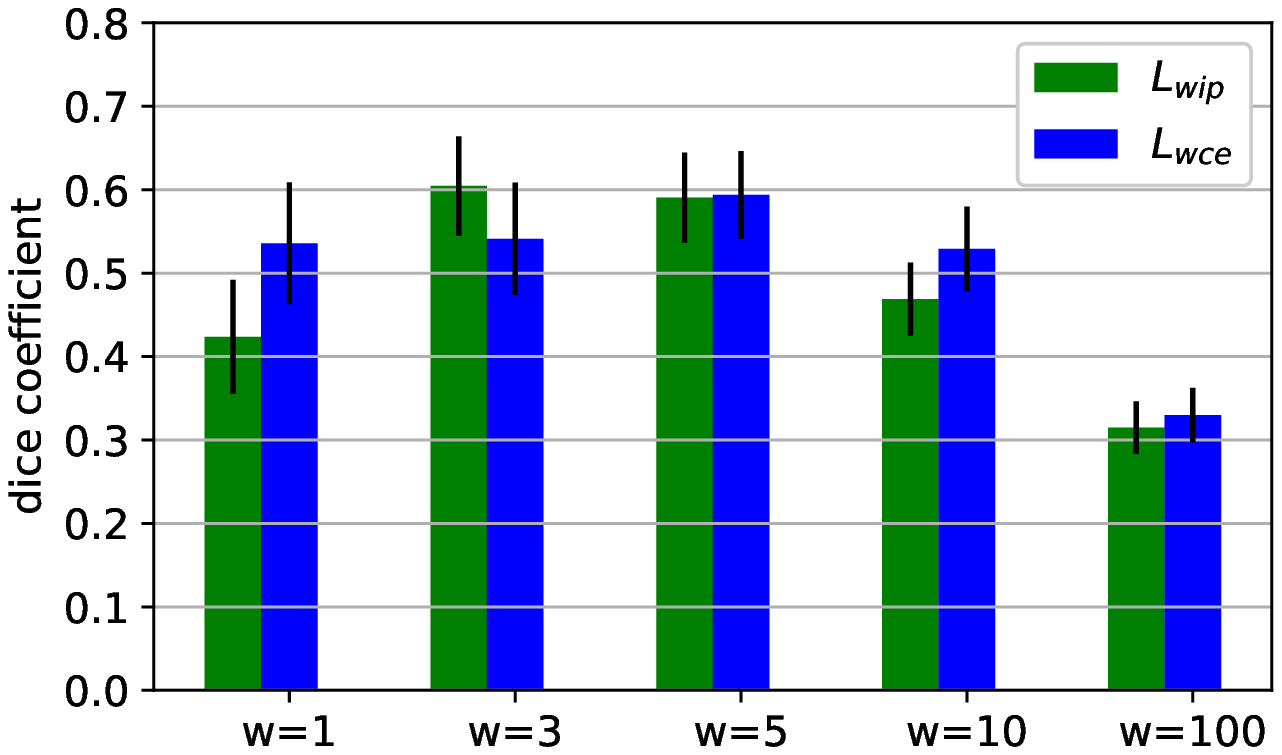}
  \caption{Test dice coefficient of FMI using $L_{wip}$ and $L_{wce}$ loss function. $W$ indicates the weight of tract.}\label{fig2}
\end{figure}

Fig.~\ref{fig1} (left) presents the test DC of FMI for different combinations of input images. The figure shows that all combinations gave similar performances. Therefore, we used the simplest and most computation-efficient input, i.e. tensor only.

The performance when varying the loss function ($L_{wip}$ and $L_{wce}$) and tract weight is provided in Fig.~\ref{fig2}. $L_{wip}$ in combination with $W=3$ gave the best result. Both loss functions had instable performance when $W>5$, especially $L_{wce}$. Based on the comparison, we used $L_{wip}$ ($W=3$) in the remainder of the experiments.

Since Fig.~\ref{fig1} (right) shows that the 3D U-Net architecture in combination with the Adam optimizer yielded a better performance than the other methods using either a V-Net architecture or the Nadam optimizer, we will adopt this combination in our method.

% \begin{figure}[!ht]
%  \centering
%   \includegraphics[scale=0.4]{loss_weights_bar_test.eps}
%   \caption{Test dice coefficient of V-Net based models using $L_{wip}$ (green) and $L_{wce}$ (blue) loss function. From left to right, $ W = [1, 3, 5, 10, 100] $.}\label{fig2}
% \end{figure}

\subsection{Validation results}
% After optimization, we trained our 3D U-Net based model using $L_{wip}$ ($W=3$) and Adam optimizer. 

% \begin{figure}[!h]
%   \centering
%   \subfigure[Axial]{
%     \includegraphics[width=0.32\textwidth]{test_fig//1_1.eps}}
%   \hspace{-0.5cm}
%   \quad
%   \subfigure[Sagittal]{
%     \includegraphics[width=0.32\textwidth]{test_fig//1_2.eps}}
%   \hspace{-0.5cm}
%   \quad
%   \subfigure[Coronal]{
%     \includegraphics[width=0.32\textwidth]{test_fig//1_3.eps}}
%   \caption{predictions of FMI and CST}\label{fig3}
% \end{figure}

% \begin{figure}[!h]
%   \centering
%   \subfigure{
%     \includegraphics[scale=0.5]{FMI_1714_2.png}}
%   \quad
%   \subfigure{
%     \includegraphics[scale=0.5]{FMI_1714_3.png}}
%   \caption{FMI segmentation: the overlap(red) comparison between reference standard (blue) and method prediction (yellow) in different views. The binary dice coefficient of this subject is 0.59.}\label{fig3}
% \end{figure}

\begin{figure}[!ht]
  \centering
  \subfigure[]{
    \includegraphics[width=0.23\textwidth]
{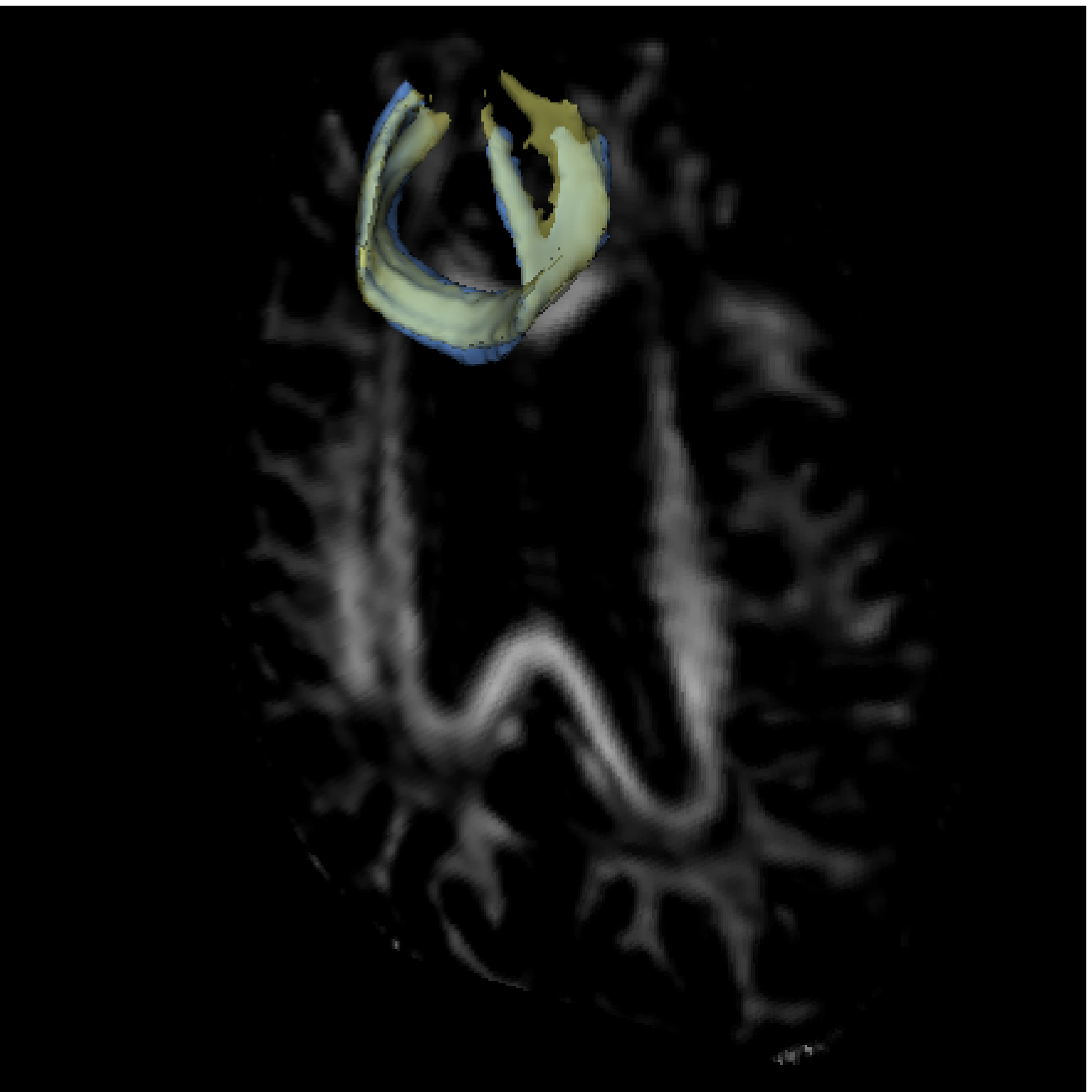}}
  \hspace{-0.5cm}
  \quad
  \subfigure[]{
    \includegraphics[width=0.23\textwidth]{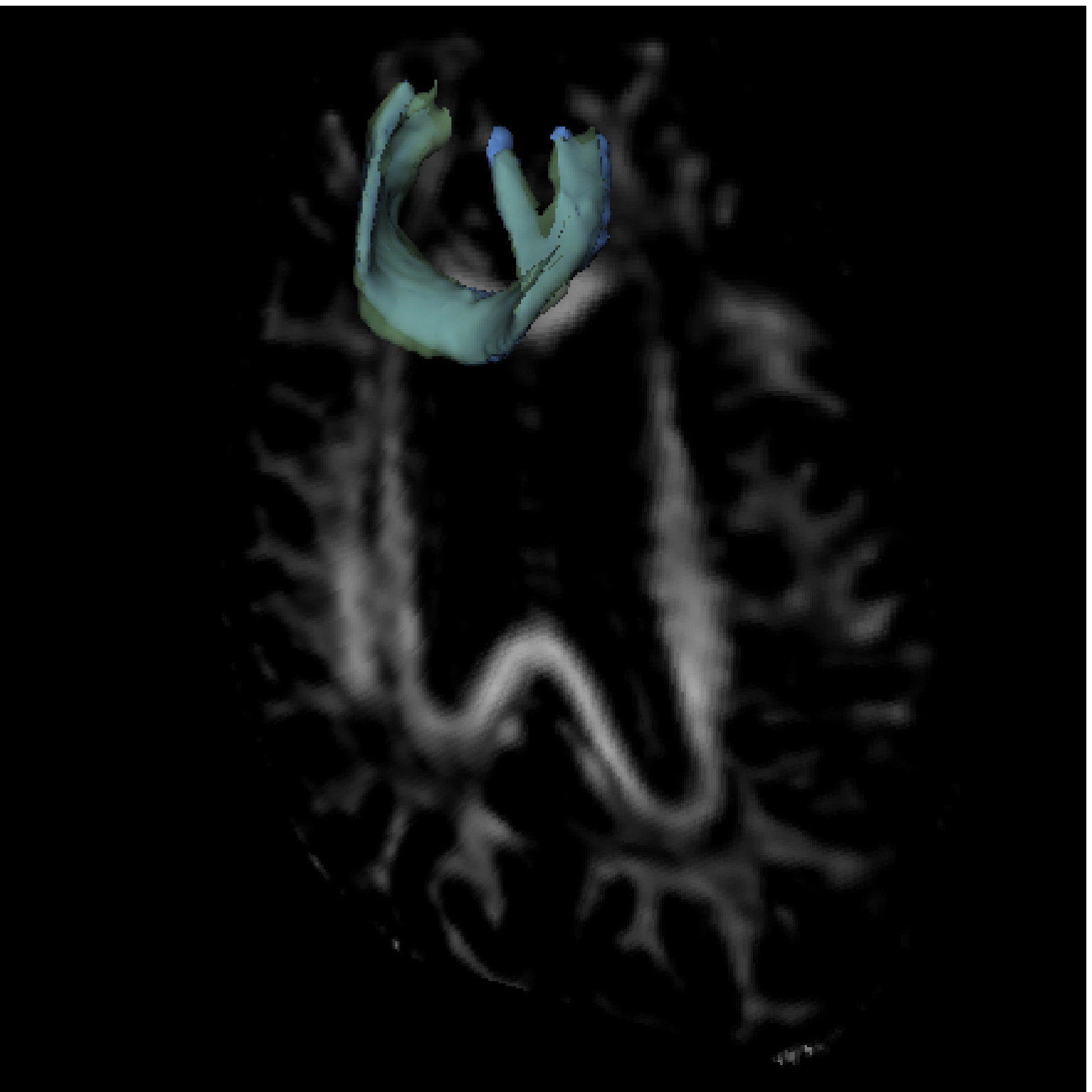}}
  \hspace{-0.5cm}
  \quad
  \subfigure[]{
    \includegraphics[width=0.23\textwidth]{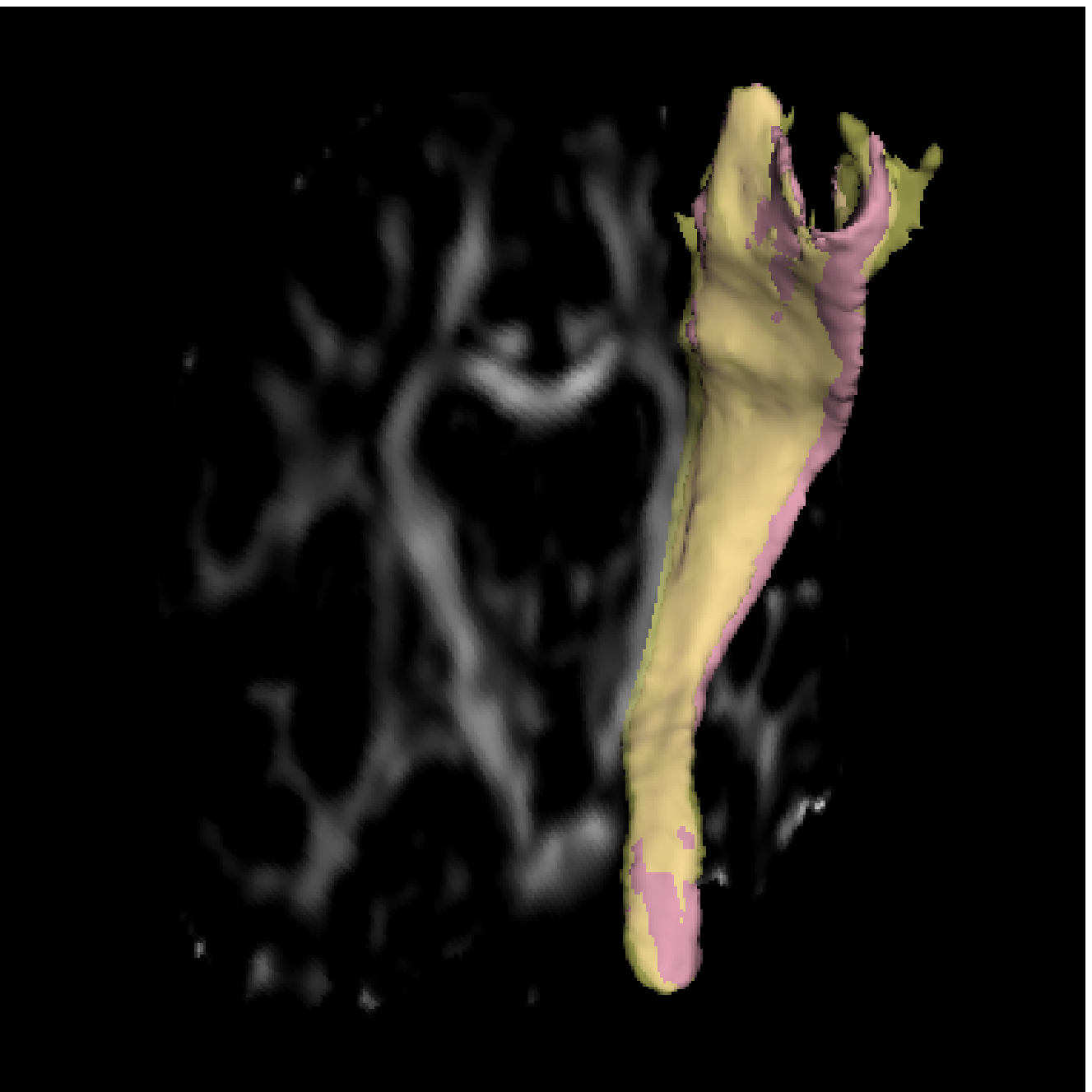}}
  \hspace{-0.5cm}
  \quad
  \subfigure[]{
    \includegraphics[width=0.23\textwidth]{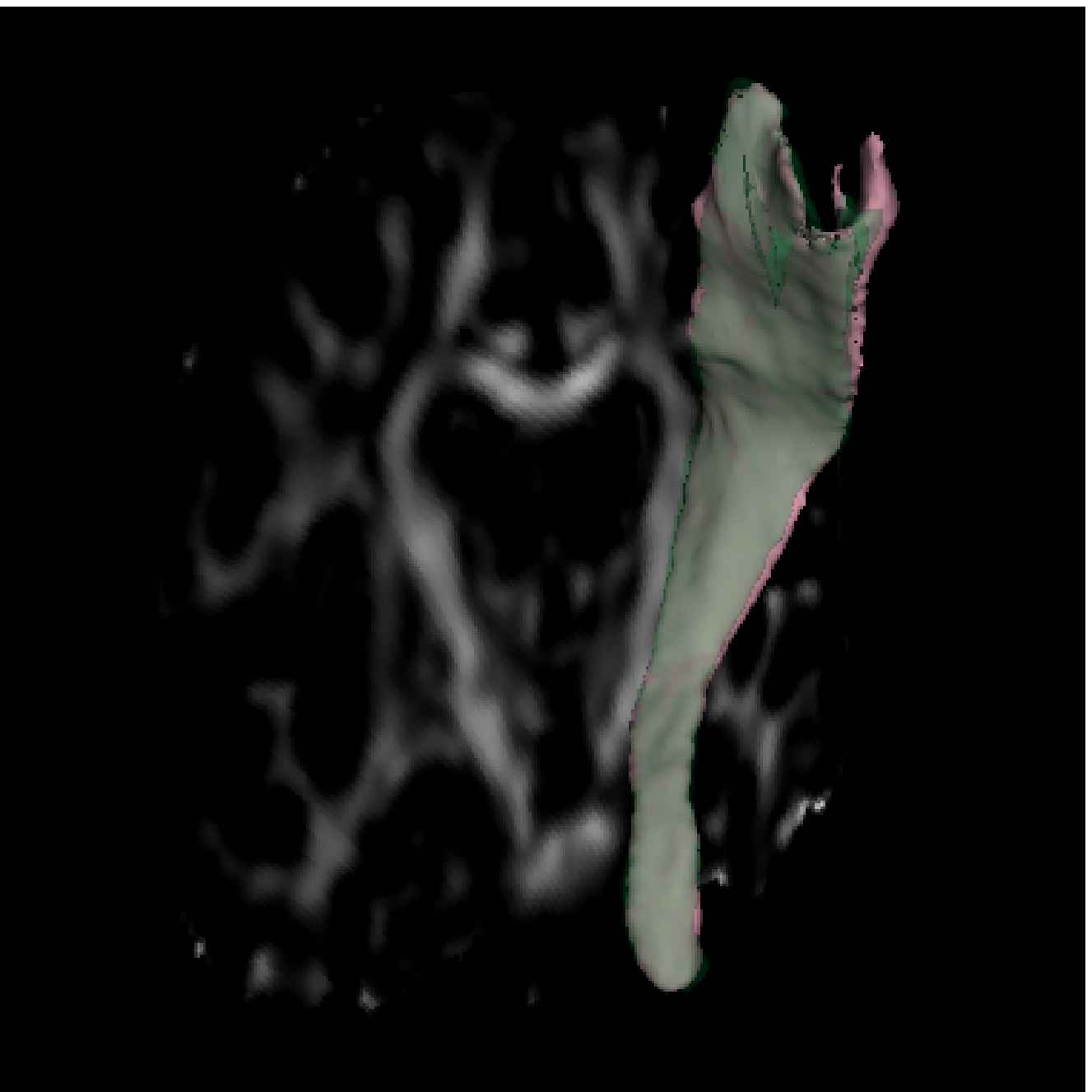}}
  \caption{Visualization of segmentation results: (a) FMI (blue) and reference (yellow), $DC=0.67$; (b) FMI of the first scan and rescan (green), $K=0.79$; (c) right CST (pink) and reference, $DC=0.76$; (d) right CST of the first scan and rescan, $K=0.84$.}\label{fig3}
    \vspace{0.2cm}
  \subfigure{
    \includegraphics[scale=0.46]{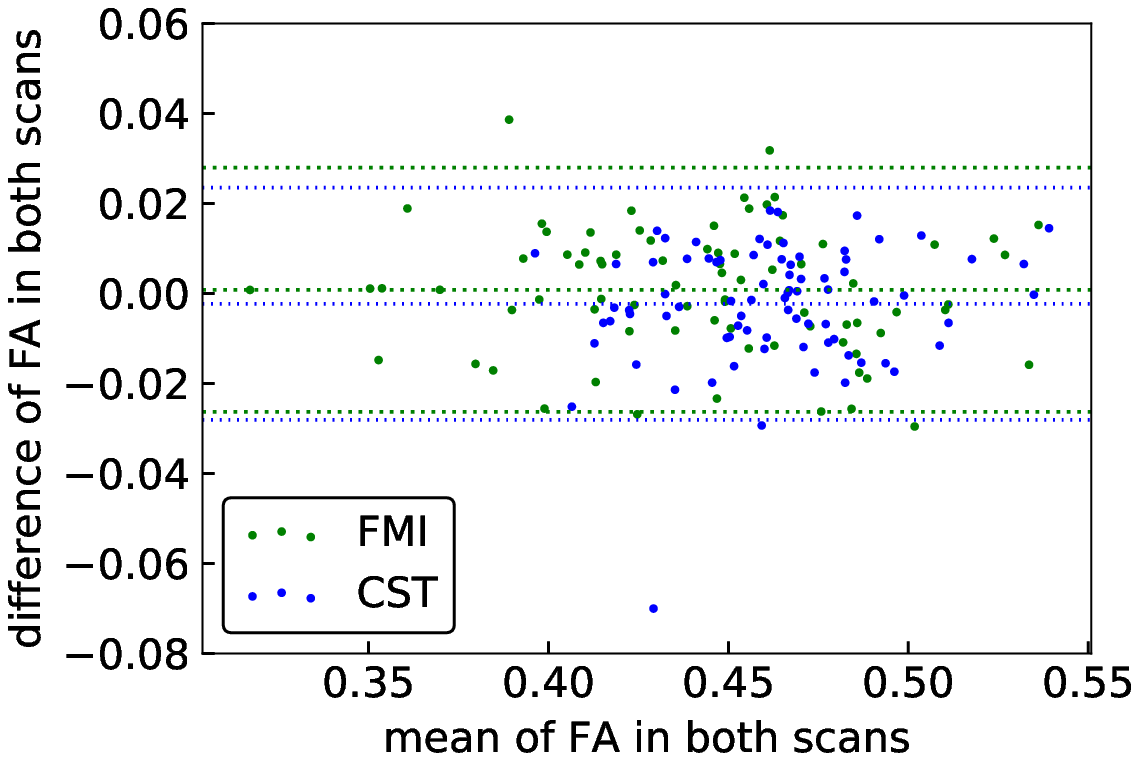}}
  \quad
  \subfigure{
    \includegraphics[scale=0.46]{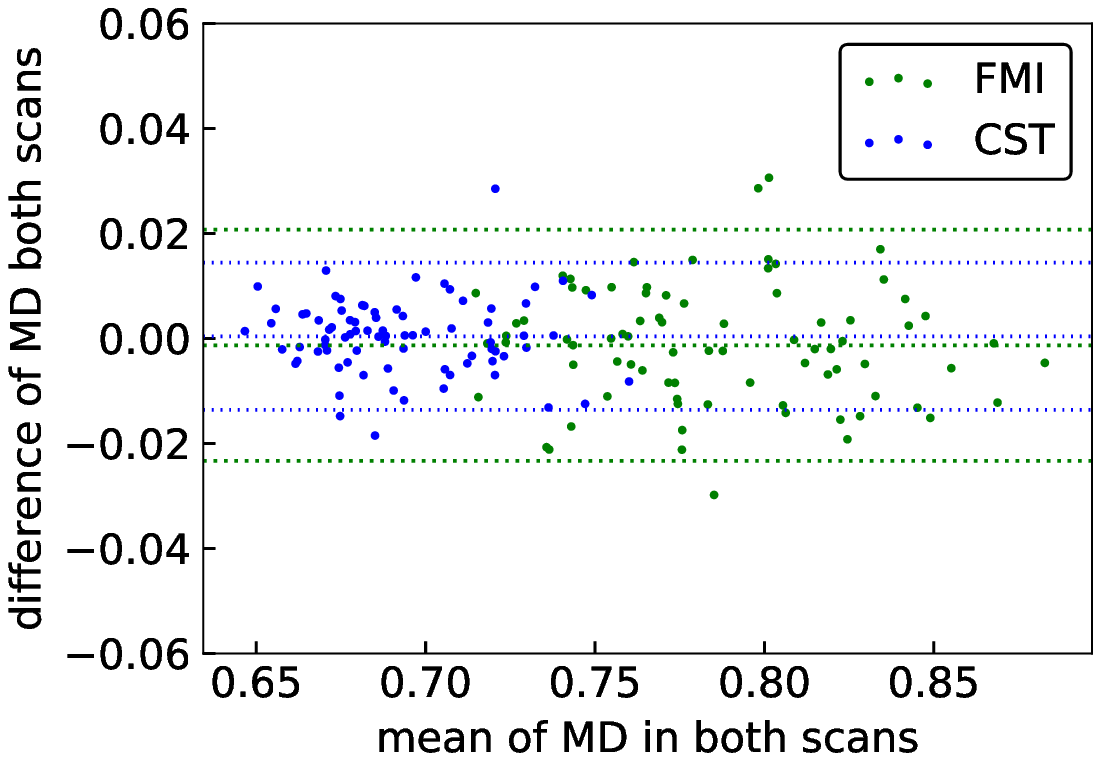}}
  \caption{The Bland-Altman plots. Difference (y-axis) and mean (x-axis) of diffusion measures (left) FA, (right) MD inside the segmented FMI and CST tract in both scans. }\label{fig4}
\end{figure}

% \begin{figure}[!h]
%   \centering
%   \subfigure{
%     \includegraphics[scale=0.5]{CST_1714_1.png}}
%   \quad
%   \subfigure{
%     \includegraphics[scale=0.5]{CST_1714_2.png}}
%   \caption{CST segmentation: the overlap(red) comparison between reference standard (blue) and method prediction (yellow) in different views. The binary dice coefficient of this subject is 0.79. }\label{fig5}
% \end{figure}
Fig.~\ref{fig3} provides a visualization of our segmentation result. It overlaps with the reference standard in (a) and (c) for FMI and right CST, respectively. The mean test DC of FMI is 0.66 (SD 0.06), that of CST is 0.77 (SD 0.03).

Fig.~\ref{fig3} (b)(d) provide its overlaps with segmented rescan, which was registered by rigidly aligning the FA images. Table~\ref{tab1} gives the reproducibility statistics. Typically, a $K>0.60$ indicates ``substantial'' agreement between raters, and a $K>0.80$ for ``almost perfect'' \cite{landis1977measurement}. Our mean $K$ for FMI longitudinal-segmentations achieved 0.74 and 0.80 for CST. The $R^2$ and $K$ show that our method has better reproducibility than reference. Moreover, there was no difference in our longitudinal-measures (FA, MD, volume, paired t-test, $p>.1$). Our mean FA and MD are consistent with that of the reference. These results show that our method is applicable in longitudinal analysis of WM microstructure. 

Fig.~\ref{fig4} provides subject-wise reproducibility in determining diffusion measures. The Bland-Altman plots show that almost all differences are within the $95\%$ limits of agreement and the mean of which is close to zero, indicating no consistent bias in longitudinal-measures. Additionally, Fig.~\ref{fig4} (right) shows that the MD is a discriminative feature for FMI and CST tract.

\begin{table*}
\scriptsize
\caption{Tract-specific reproducibility statistics. 
MD x $10^{-3} mm^{2}/s$. The ``ref'' indicates reference standard; ``prop'' indicates proposed method; SD standard deviation; ``diff (SD)'' indicates averaged absolute-differences between both scans; ``mean (SD)'' indicates mean value over all scans; $R^2$ is the $R^2$ value of OLS regression for measures in both scans; ``Vol'' indicates tract volume in ml; ``$K$'' indicates the Cohen's kappa; $^{*}$ significantly ($99\%-CI$) improved from the reference, t-test, $p<.01$.}\label{tab1}
\begin{tabular*}{1.0\textwidth}
{@{}@{\extracolsep{\fill}}cccccccccccccc@{}}
\hline
\multicolumn{1}{c}{ }  &
\multicolumn{3}{c}{FA} &
\multicolumn{1}{c}{}   &
\multicolumn{3}{c}{MD} &
\multicolumn{1}{c}{}   &
\multicolumn{3}{c}{Vol}  &
\multicolumn{1}{c}{}   &
\multicolumn{1}{c}{$K$} \\
\cline{2-4}\cline{6-8}\cline{10-12}\cline{14-14}
 &diff&mean&$R^2$&  &diff&mean&$R^2$&  &diff&mean &$R^2$ \\
 \hline
{\itshape FMI}\\
{ref}    &.012 (.009) &.44 (.04) &.89 & &.0082 (.007) &.79 (.04)  &.93 & &.26 (.21)& 3.3 (.53)&.66& &.64 (.02)\\
{prop}     &.011 (.008) &.44 (.05) &.91 & &.0089 (.007)&.79  (.04) &.93 & & .23 (.16) &3.8 (.58)&.79& &$.74^{*}$ (.01)\\
\hline
{\itshape CST}\\
{ref}    &.011 (.008) &.46 (.03) &.83 & &.0053 (.005) &.70 (.03) &.92  &  &.64 (.50)& 6.1 (.93)&.39& &.72 (.04)\\
{prop}     &.009 (.003)&.46 (.03) &.84 & &.0052 (.004)&.69 (.03)  &.93 & &$.41^{*}$(.26) &6.5 (.69)&.52& &$.80^{*}$ (.07)\\
\hline
\end{tabular*}
\end{table*}

% \begin{figure}[!ht]
%   \centering
%   \subfigure{
%     \includegraphics[scale=0.5]{bland_plot_FA_CST_80pairs.eps}}
%   \quad
%   \subfigure{
%     \includegraphics[scale=0.5]{bland_plot_MD_CST_80pairs.eps}}
%   \caption{The Bland-Altman plots. Difference (y-axis) and mean (x-axis) of diffusion metrics (a) FA, (b) MD inside the segmented FMI and CST tract in both scans. }\label{fig4}
% \end{figure}

% We developed and evaluated a novel deep learning method for direct WM tract segmentation. The method was trained and applied on a large set of low resolution DTI images and showed very good reproducibility, higher than the reference standard. Therefor it can be applied to longitudinal imaging studies to investigate the process of neurodegeneration in WM microstructure as can be assessed with diffusion MRI. 
\section{Discussion}
We developed and evaluated a novel deep learning method for direct WM tract segmentation. The method was trained and applied on a large set of low resolution DTI images and showed very good reproducibility. Therefore it can be applied to longitudinal imaging studies to investigate the process of neurodegeneration in WM microstructure as can be assessed with diffusion MRI. 

Strengths of this study are the large size of dataset, which is representative of clinical variation, and the reproducibility validation in both segmentation and determining diffusion measures. Reproducibility is an essential indicator of a method that can be applied in clinical practice to ensure reliable and reproducible results. Moreover, comparing with the tractography-based methods, our direct method enables to segment a 3D tract in 0.5 seconds, and therefore avoid the processing time and storage space of tractography for researchers who only focus on the analysis of diffusion measures. 

% The age range, amount, independence and low resolution of $D2_{test}$ is representative of the clinical variability.
% To the best of our knowledge, this is also the first comparison between 3D U-Net and V-Net architecture.
% Moreover, our feature selection results provide a reference for other researchers attempting to segment WM tract with DTI features.

% Based on the results of FMI segmentation, we found that the mean DC of method ($D1b_{train}$) decreased from 0.68 ($D1b_{test}$) to 0.61 ($D2_{test}$) when tested on a larger and train-independent dataset. This is acceptable, since the subjects in $D1b_{train}$ is much older than those in $D2_{test}$. On the other hand, the DC tested on independent $D2_{test}$ increased from 0.61 ($D1b_{train}$) to 0.66 ($D2_{train}$) by enlarging the training dataset. The increase is lower than we expected, we suspect that the great diversity of training dataset increases the robustness and difficulty of learning at the same time.

Based on the results of FMI segmentation we concluded that both the dependency of train and test datasets and their respective sizes are important for the resulting performance. If a much older training ($D1b_{train}$) than testing ($D2_{test}$) dataset is used, performance is suboptimal ($DC=0.61$, $D1b_{train}$/$D2_{test}$ vs. $DC=0.68$, $D1b_{train}$/$D1b_{test}$). On the other hand, a large, diverse and test-independent training dataset increases the robustness and difficulty of learning at the same time ($DC=0.66$, $D2_{train}$/$D2_{test}$).

% The DC of the right CST segmented by our method (DC=0.77) is lower than that reported in \cite{6} (DC=0.83). which provides more information of crossing-fiber.The main differences between two works are: their smoothly annotated reference, higher-order input derived from 7 T scanner and small test set of only 5 subjects.The reasons we did not do direct comparison are that our DC was computed with unsmooth reference standard as Fig.~\ref{fig3} shows, and that our large test dataset of 1036 subjects from the train-independent cohort.

The paper by Wasserthal et al. \cite{wasserthal2017direct} is the only published deep learning method of WM tract segmentation that we are aware of. Our test DC of the right CST ($DC=0.77$) is lower than that reported in \cite{wasserthal2017direct} ($DC=0.83$). The main differences between two works are: they takes high resolution (7T) based input and semi-manual annotated reference, stacks four 2D models and is tested on only 5 subjects; while ours is applicable for a low-resolution dataset (1.5T), uses one 3D model and tested on a train-independent cohort of 1036 subjects. We suspect that the differences in the quality of the reference standard and the data are the main causes of this performance difference.

% A limitation of our method is that we take a single tract ROI. This is mainly because of the large whole input size of 210$\times$211$\times$123$\times$6 and the computation limitation of 3D convolution which used for preserving the continuity of tract. Another limitation is our low quality reference standard. Validation is difficult because semi-manual annotation can not be obtained for such a large dataset.

A limitation of our method is that we take a single tract ROI. This is mainly because of the large whole input size of 210$\times$211$\times$123$\times$6 and the limitation of 3D convolution, which used for preserving the continuity of the tract. Another limitation is our low quality reference standard. Validation is difficult since the semi-manual annotation can not be obtained for such a large dataset.

For future work, our method will be applied in a dementia population. We will tackle the computation limitation of taking whole brain volume as input.

We conclude that our direct WM tract segmentation method has very good reproducibility and comparable performance to the reference standard. This is the first deep learning based method of WM tract segmentation developed on such a large-scale dataset. Our method can lead toward a faster, more lightweight way of diffusion measures analysis, thereby, reducing the time-consuming of segmentation, the complexity of pipeline setting and the required storage space.
% The method may be used in clinical low-resolution settings and is applicable in longitudinal analysis of neurodegeneration.
%
% the environments 'definition', 'lemma', 'proposition', 'corollary',
% 'remark', and 'example' are defined in the LLNCS documentclass as well.
%
%
% ---- Bibliography ----
%
% BibTeX users should specify bibliography style 'splncs04'.
% References will then be sorted and formatted in the correct style.
%
\bibliographystyle{splncs04_use}
\bibliography{WM_tract_segmentation_use}
\end{document}